\documentclass[slac_one]{revtex4}
\usepackage{graphicx}
\usepackage{fancyhdr}
\begin{document}

\title{Recent Results from Super-Kamiokande} 

%

\author{H. Sekiya (for the Super-Kamiokande Collaboration)}
\affiliation{Kamioka Observatory, Institute for Cosmic Ray Research, University of Tokyo\\
456 Higashi-Mozumi, Kamioka, Hida, Gifu, 506-1205 JAPAN}

\begin{abstract}
The recent results on the oscillation analyses of solar neutrino and the
 atmospheric neutrino measurements in Super-Kamiokande are presented. 
Recent status of the detector is also reported.
\end{abstract}

\maketitle

\thispagestyle{fancy}


\section{INTRODUCTION} 
Super-Kamiokande (SK) is a water Cherenkov detector containing 50,000 
tons of pure water. It is located 1000 meters underground 
(2,700 meters of water equivalent) in Kamioka zinc mine in Japan. 
The detector consists of a main inner detector and an outer veto
detector. Both detectors are contained within a cylindrical 
stainless steel tank 39.3 m in diameter $\times$ 41.4 m in height. 
The usual fiducial mass for neutrino measurements is 22.5 ktons 
with boundaries 2.0 m from the inner surface. 

The first phase of the SK (SK-I) started in April, 1996,
and terminated in July, 2001. A total of 11,146 PMTs with 
20-inch diameter photocathodes provided active light 
collection over 40$\%$ of the surface of the inner detector.
Then, in spite of the loss of numerous PMTs in an accident,
the second phase (SK-II) started in December, 2002. 
 A total of 5,182 20-inch PMTs, each protected by acrylic 
and fiber-reinforced plastic (FRP) cases, were mounted on 
the inner detector, providing 19$\%$ photocathode coverage 
during this period. SK-II ran until October 2005. 
In the analysis of the SK-II data, analysis methods had to be 
revised due to the loss of detector sensitivity, however
it turned out that the revised method also effective in 
the SK-I data analysis.

In July 2006, SK detector was totally recovered with 40$\%$ 
photocathode coverage and started taking data as SK-III.
During this phase, the water flow in the tank had been tuned up 
and the low energy background was reduced.
In order to further enhance its performance, SK-III was terminated in
August 2008, and the readout electronics and online data acquisition 
system were upgraded. The fourth phase (SK-IV) started in September 2008.

In this paper we report updated SK-I + SK-II results
of solar and atmospheric neutrino oscillation analyses
and the detector status of SK-III and SK-IV.

\section{SOLAR NEUTRINO (SK-I and SK-II)}
The advantages of SK in the solar neutrino observation are the
time variation measurement, the direction-sensitivity,
and the very precise measurement of its spectrum due to the well 
calibrated energy of recoil electrons\cite{nakahata}.

\begin{figure}[htbp]
\begin{center}
\includegraphics[scale=0.22]{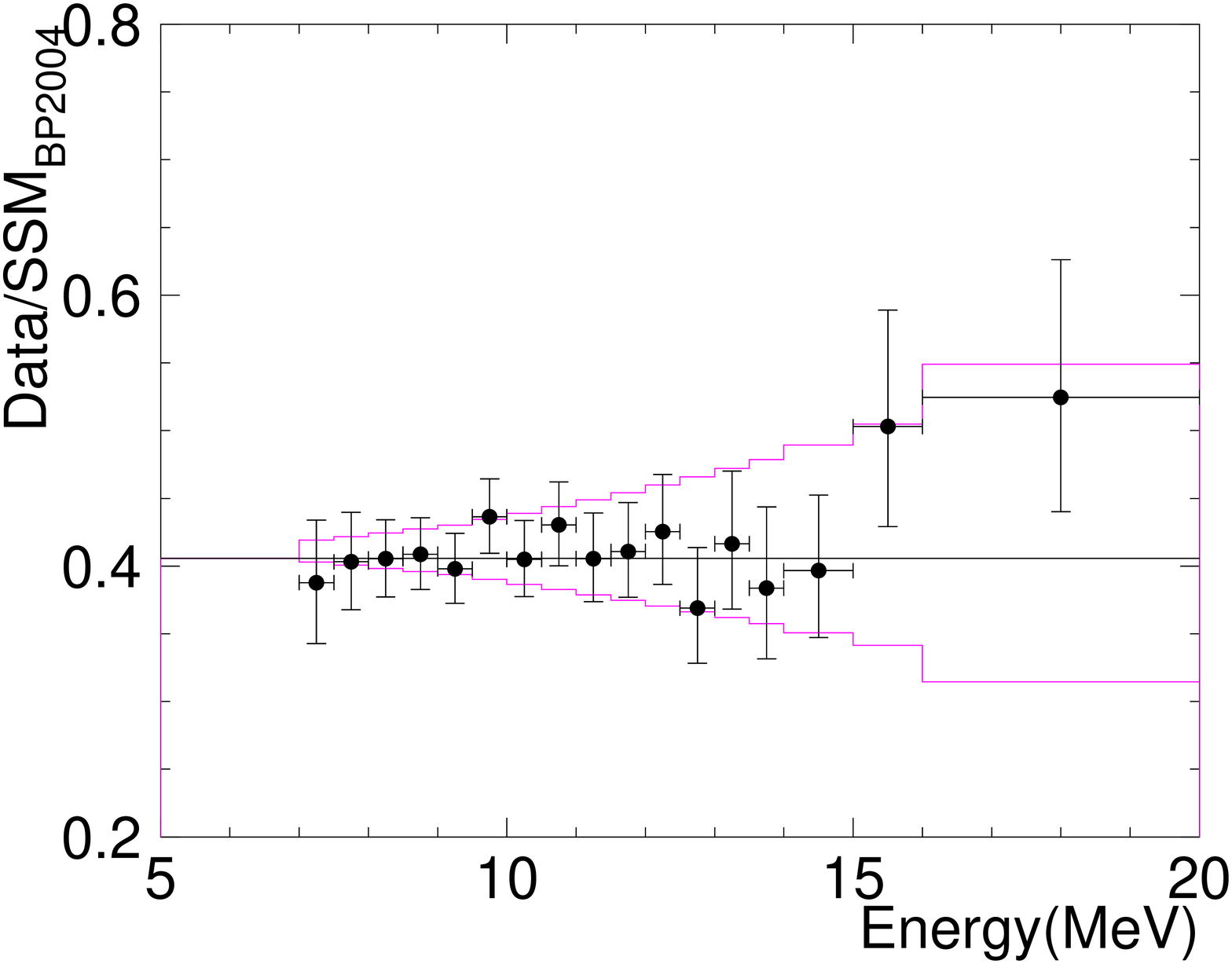}
\caption{Ratio of observed and expected energy spectra.  The purple 
lines represent a $\pm1$ sigma level of the energy correlated 
systematic errors.  The black line represents the SK-I 1496-day average 
and shows agreement with SK-II.}
\label{fig:energy_spectrum}
\end{center}
\end{figure}

The solar neutrino flux can be derived from the extracted number of
signal events by fitting signal$+$background shapes to the recoil angle
distribution relative to the Sun ($\cos\theta_{\rm Sun}$).
The observed solar neutrino flux in each phase of SK is summarized in
Table~\\ref{tab:flux}.
The SK-II flux value is statistically consistent with the SK-I value.
Figure~\ref{fig:energy_spectrum} shows the SK-II observed energy spectrum 
divided by the expected spectrum without oscillation determined 
from the BP2004 SSM\cite{bp2004}.
The line through the spectrum represents the total SK-I average.
SK-II shows excellent agreement with SK-I.

\begin{table}[htbp]
\begin{center}
\begin{tabular}{l|r|r|l|l}
\hline
Phase  & Live time(days) & Energy Range(MeV) & Number of signals & Flux ($\times 10^6 {\rm cm}^{-2} {\rm sec}^{-1}$) \\ \hline \hline
SK-II  & 791 & 7.0-20.0 & $7212.8^{+152.9}_{-150.9}$(stat)$^{+483.3}_{-461.6}$(sys) &
$2.38\pm 0.05$(stat)$^{+0.16}_{-0.15}$(sys) \\ \hline
SK-I  & 1496 & 5.0-20.0 & $22404\pm 226$(stat)$^{+784}_{-717}$(sys) &
$2.35\pm 0.02$(stat)$\pm 0.08$(sys) \\ \hline
\end{tabular}
\caption{Observed solar neutrino flux in SK-I and SK-II}  
\label{tab:flux}
\end{center}
\end{table}

\begin{figure}[htbp]
\begin{center}
\includegraphics[scale=0.27]{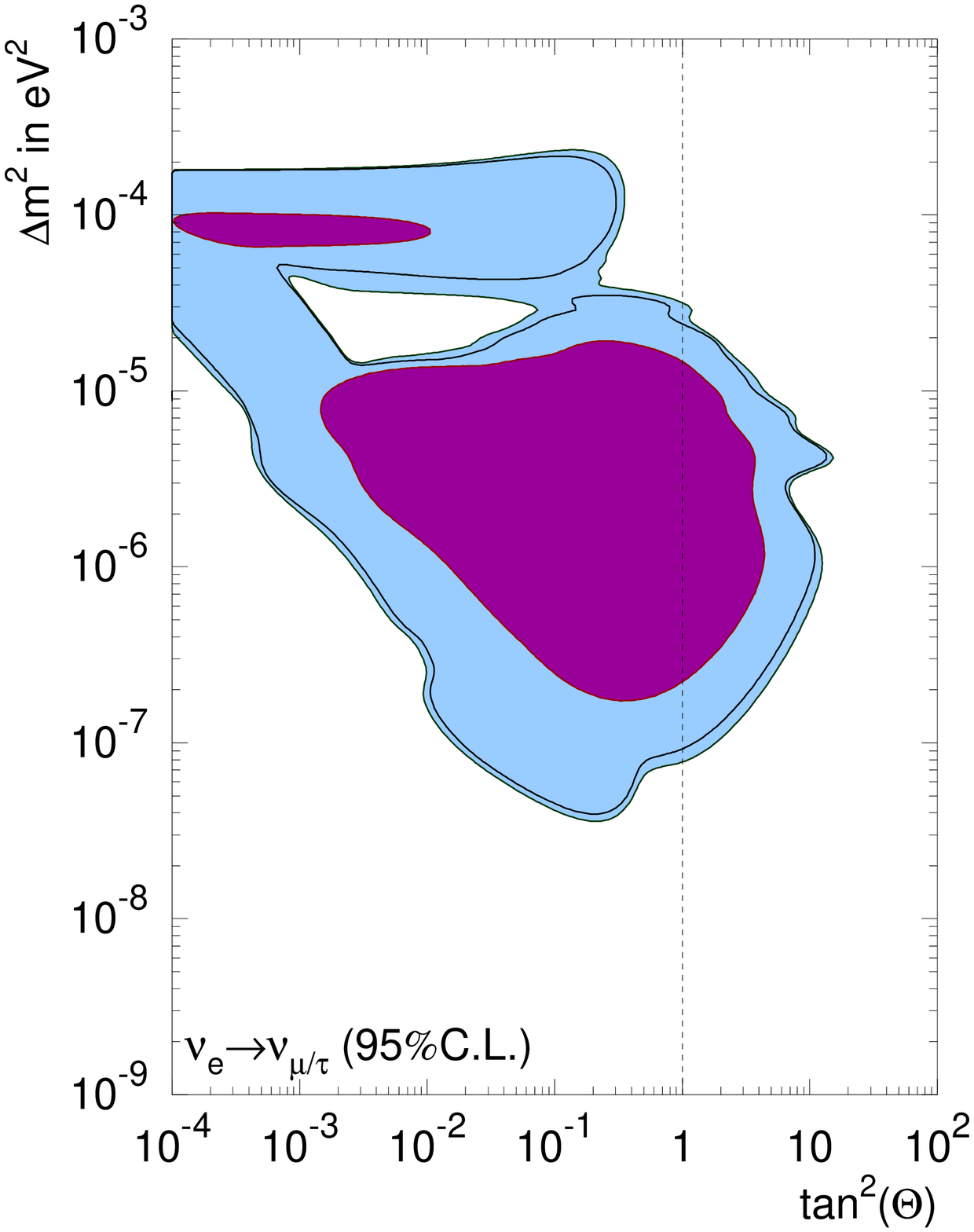}
\includegraphics[scale=0.27]{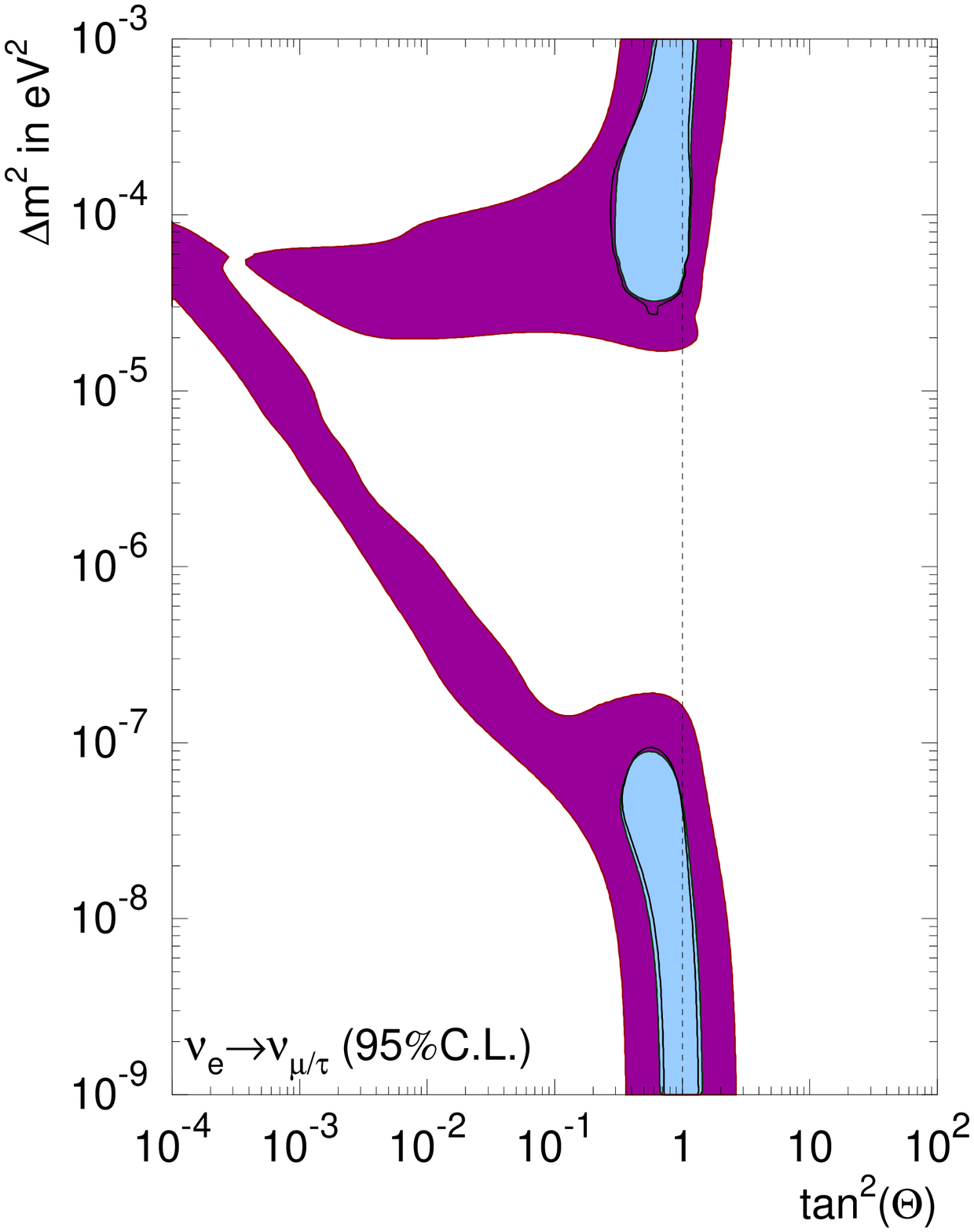}
\includegraphics[scale=0.27]{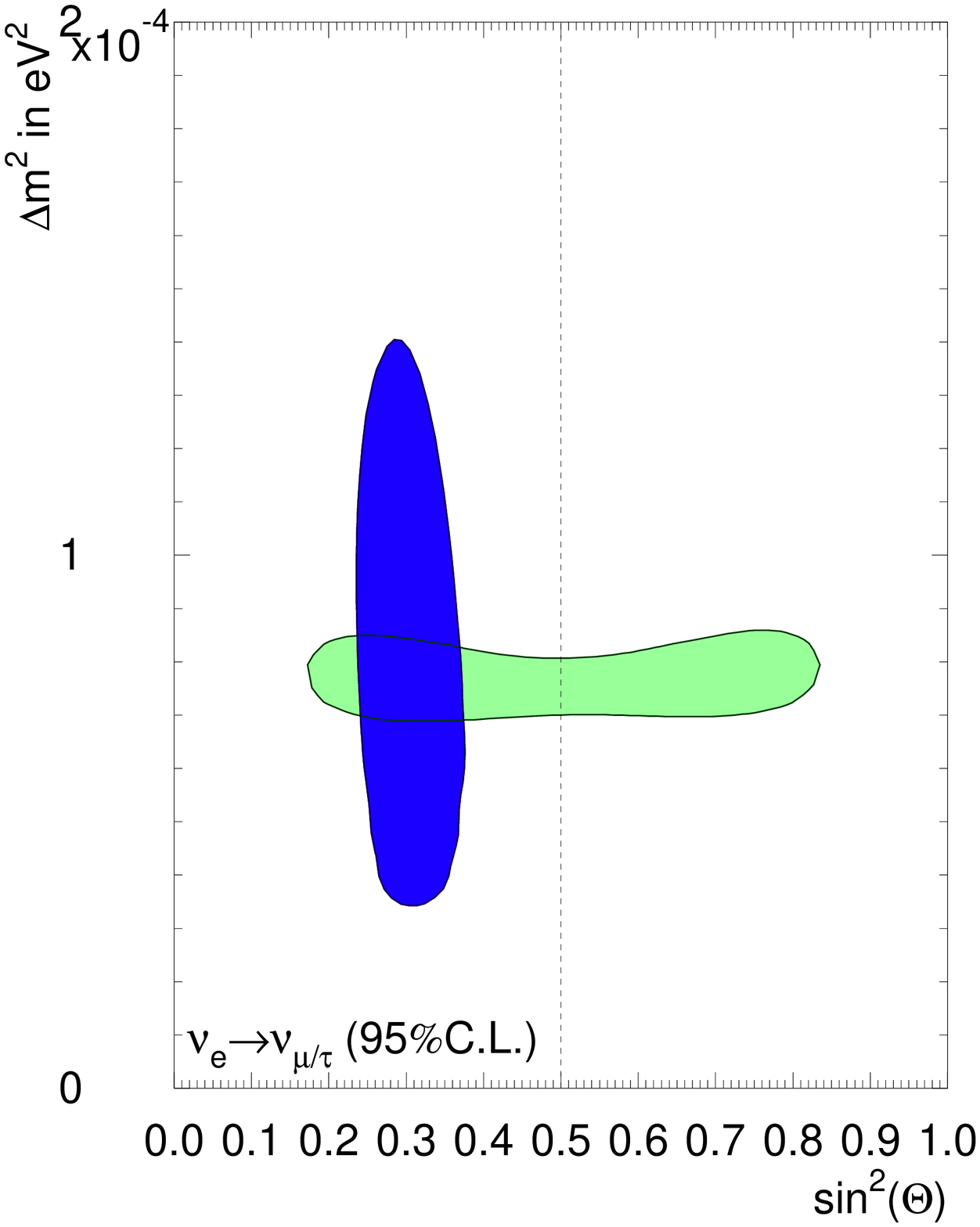}
\caption{The left plot shows SK excluded areas.  The purple region is SK-II and the light blue region is SK-I with SK-II.  The black line shows SK-I only, and evidence of increased exclusion can be seen with the addition of SK-II data.  The center plot shows SK allowed regions (same representative colors as the exclusion contours) with the $^8$B flux constrained to the SNO total flux measurement.  The hep flux is a free parameter.  The right plot shows the SK-I and SK-II combined contour with SNO and radiochemical solar experimental data (blue contour).  The green contour is the KamLAND\cite{kamland} electron anti-neutrino oscillation result.}
\label{fig:contours}
\end{center}
\end{figure}

The determination of the solar neutrino oscillation parameters 
in SK-I and SK-II is accomplished in the same way\cite{parker}.
Two neutrino oscillation is assumed and for each set of
oscillation parameters, a $\chi^2$ minimization of the total $^8$B
and hep neutrino flux is fit to the data.
This yields exclude regions, while by constraining the $^8$B flux
to the total NC flux value from SNO\cite{sno}, allowed regions
can be obtained.
Figure~\ref{fig:contours} shows both excluded and allowed regions
at 95$\%$ confidence level.

The combination of other solar neutrino experiments such as SNO and 
radiochemical results (Homestake, GALLEX and  SAGE) with the SK combined
analysis is also accomplished and shown Figure~\ref{fig:contours}.
The best fit parameter set is tan$^2\theta=0.40$ and $\Delta m^2=6.03\times
10^{-5}$eV$^{2}$.

\section{ATMOSPHERIC NEUTRINO (SK-I and SK-II)}
Super-Kamiokande has reported that the atmospheric neutrino data 
are well consistent with the pure $\nu_\mu \leftrightarrow \nu_\tau$ two 
flavor oscillation scheme\cite{Ashie}.
Recently $\nu_\mu \leftrightarrow \nu_\tau$ oscillation analyses for zenith angle
distributions and L/E distribution are re-performed 
using SK-I and SK-II data with improved analysis methods 
to get more stringent constraint on the oscillation parameters.
The changes to the simulation include: an update of the atmospheric
neutrino flux model to the ``Honda06'' model\cite{honda}; 
various modifications to the neutrino interaction model, NEUT\cite{neut}
({\it e.g.}, change
of quasi-elastic and single pion axial mass to $M_A$ = 1.2 GeV, addition of lepton mass effects
for charged-current single pion production, and addition of the
pion-less delta decay channel $\Delta \rightarrow N\gamma$);
improvements to the detector simulation model of light reflections and scattering;
better tuning of outer detector parameters in the simulation; and improvements in the ring
counting algorithm. Additionally, at the time of re-analysis, the systematic uncertainties were
re-evaluated and a few new uncertainties were added.

\begin{figure}[htbp]
\begin{center}
\includegraphics[scale=0.4]{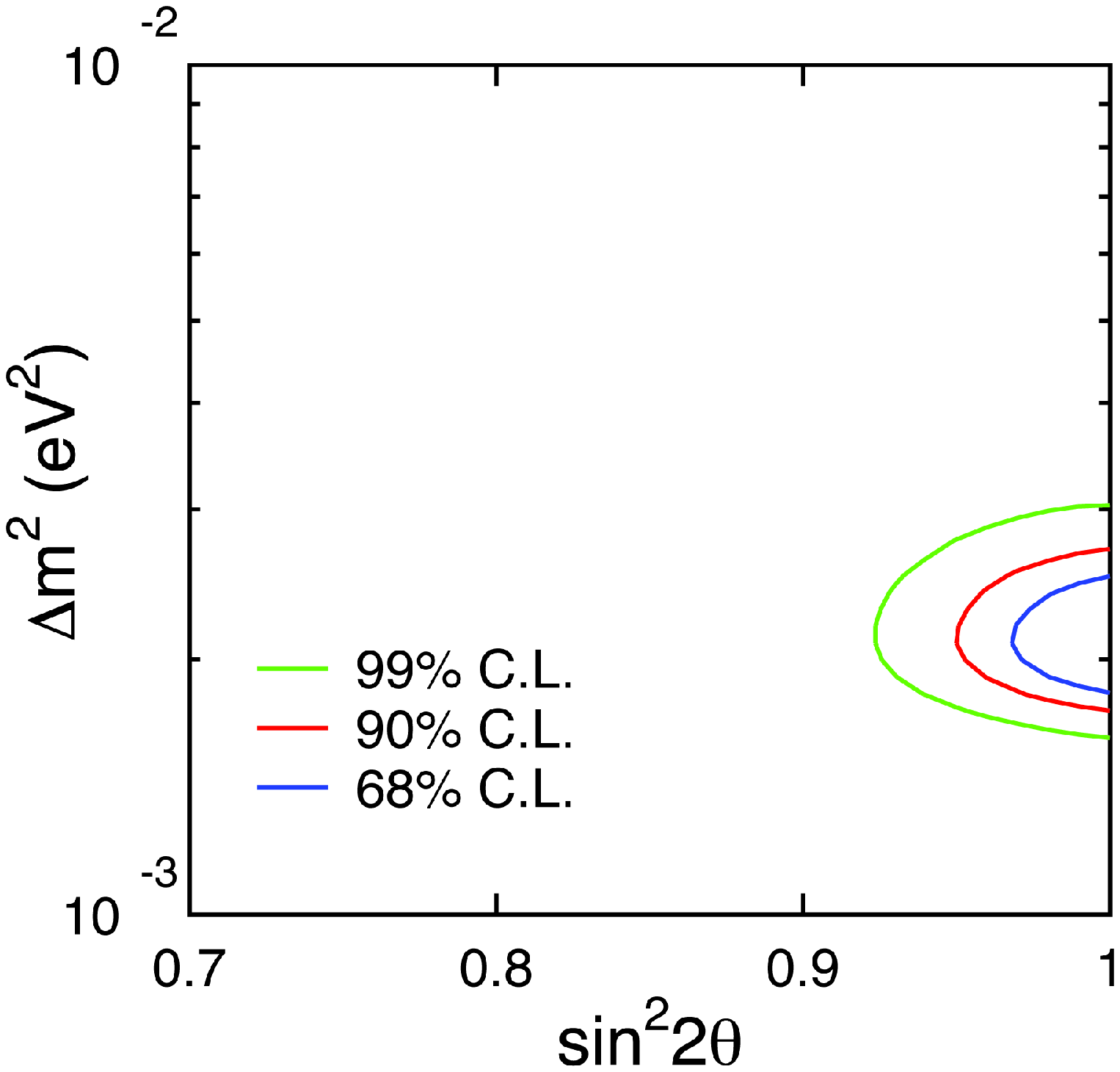}
\includegraphics[scale=0.4]{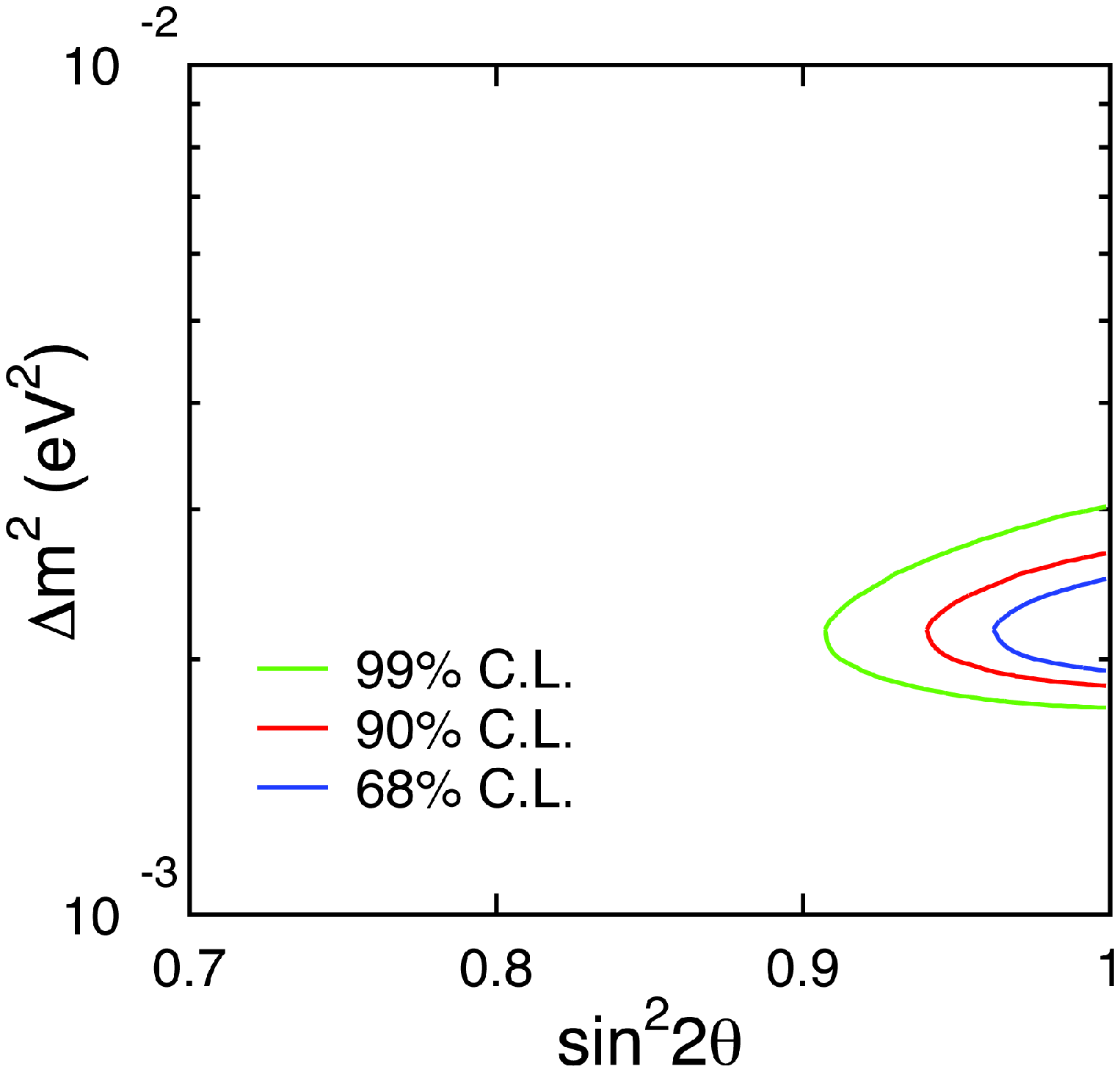}
\caption{The left plot shows allowed oscillation parameters by the
 zenith angle analysis and the right plot shows the ones by the  L/E analysis.}
\label{fig:atm_contours}
\end{center}
\end{figure}

Atmospheric neutrino data are categorized into fully-contained (FC),
partially-contained (PC), and upward-going muon (UP$\mu$) events.
In the zenith angle analysis, 1489 days of SK-I FC/PC data
1646 days of UP$\mu$ SK-I data, 799 days of SK-II FC/PC data and
828 days of UP$\mu$ SK-II data are used.
All samples are divided in 10 zenith angle bins.
The definition of the event bins are same as SK-I and SK-II, and
400 bins for SK-I and 350 bins for SK-II are used in the analysis.
The number of observed events in each of 750 bins is compared with 
the Monte Carlo expectation.
A $\chi^2$ value is defined according to the Poisson probability distribution.
In the fitting, the expected number of events in each bin is recalculated 
to account for 90 systematic errors, which come from the uncertainty of 
the neutrino flux model, neutrino cross-section model, event selection, 
and the detector response.
Figure~\ref{fig:atm_contours} shows contours of allowed parameter 
regions.
Best fit parameter set is $\sin^22\theta=1.02$ and 
$\Delta m^2=2.1\times10^{-3} {\rm eV}^2$.

In the L/E analysis, more strict selection criteria are applied to the subsample data, 
because a good resolution for neutrino flight pathlength, L, and energy,
E, is required when trying to observe the expected dip in the L/E
spectrum due to oscillations.
1468 days of SK-I FC/PC $\mu$-like events and 799 days of SK-II FC/PC
$\mu$-like events are used and divided in 43 
bins of $\log_{10}$(L/E (\small{\rm km/GeV})).
The results of a minimum $\chi^2$ fit to the 43 bins with 29 systematic
error terms are shown in Figure~\ref{fig:atm_contours}.
Best fit parameter set is $\sin^22\theta=1.04$ and 
$\Delta m^2=2.2\times10^{-3} {\rm eV}^2$.

\section{STATUS OF THE DETECTOR (SK-III and SK-IV)}
\begin{figure}[htbp]
\begin{center}
\includegraphics[scale=0.9]{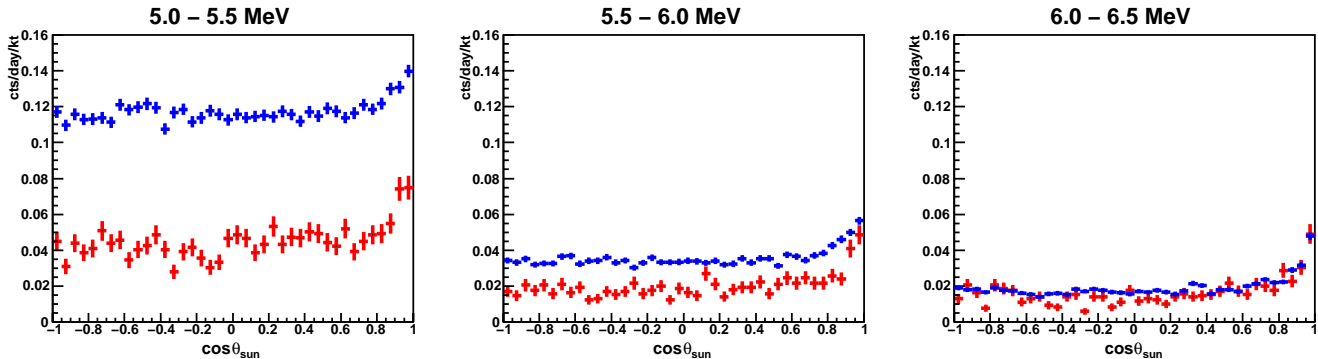}
\caption{The electron recoil angle distribution relative to the Sun in
 the central region of SK tank (16$>$z$>$-7.5m, r$<$11m). 
 The blue line shows SK-I and the red line shows SK-III.}
\label{fig:bg}
\end{center}
\end{figure}

The lower energy threshold of SK-I had been limited to around 5.0 MeV 
due to residual radon in the water emanated from the PMT/FRP, 
therefore the water purification system was upgraded 
and the water flow in SK tank had been tuned in the SK-III phase.
Figure~\ref{fig:bg} shows $\cos\theta_{\rm Sun}$ distribution of the central region.
In the 5.0-5.5MeV region, background event rate of SK-III is 
about 1/3 of that of SK-I, while the signal rate looks similar. 
Based on this background reduction, 
the trigger threshold had been lowered since April 2008,
and 100$\%$ trigger effifiency at 4.5MeV was achieved in the 
last period of SK-III.

In order to further enhance its performance, new electronics
system using custom ASIC had been developed and was installed to
SK. In the new system, all PMT hits are sent to the front-end DAQ
computers by a periodic timimg signal and software triggers are spplied
to select interesting event windows.
SK has started taking data of every hit without any hardware trigger
as ``SK-IV'' since September 2008.
This enables higher speed data taking, much lower energy threshold and
wider dynamic range 
with aiming at 
increasing detection efficiency of super nova burst neutrino,  
detecting the spectrum distortion of low energy $^8$B solar neutrino,
and  improving the energy resolution of multi-GeV atmospheric neutrino.

\section{CONCLUSION}
Super-Kamiokande (SK-I,II,III) have been operated successfully, and
more than 10 years of dataset for solar and atmospheric neutrino
are accumulated.
With the combined SK-I and SK-II dataset,
the best fit oscillation parameter set is found at tan$^2\theta=0.40$ and $\Delta m^2=6.03\times
10^{-5}$eV$^{2}$ in the solar global analysis and 
at $\sin^22\theta=1.02$ and $\Delta m^2=2.1\times10^{-3} {\rm eV}^2$ in 
the atmospheric zenith angle analysis.
In SK-III, the the background rate in the low energy region
was reduced and SK-IV has just started since September 2008 with
upgraded electronics. 
Super-Kamiokande will continue to observe every predicted effect and measure
mixing angles. In addition, high-statistics data sets make it feasible
to search for sub-dominat, exotic, and non-oscillation physics.


\end{document}